# Social Media in the Global South: A Network Dataset of the Malian Twittersphere

**Daniel Thilo Schroeder[1], Mirjam de Bruijn[2], Luca Bruls[3*] Mulatu Alemayehu Moges[4], Samba Dialimpa Badji[5], Noémie Fritz[6], Modibo Galy Cissé[7], Johannes Langguth[8], Bruce Mutsvairo[9], Kristin Skare Orgeret[10]**

1 SINTEF, Norway

2 Leiden University, The Netherlands

3 Leiden University, The Netherlands

4 OsloMet, Norway

5 OsloMet, Norway

6 University of Oslo, Norway

7 Voice4Thought, Leiden University, Mali

8 Simula, Norway

9 Utrecht University, The Netherlands

10 OsloMet, Norway

*Corresponding author: Luca Bruls l.r.bruls@asc.leidenuniv.nl

**Abstract**
With the expansion of mobile communications infrastructure, social media usage in the Global South is surging. Compared to the Global North, populations of the Global South have had less prior experience with social media from stationary computers and wired Internet. Many countries are experiencing violent conflicts that have a profound effect on their societies. As a result, social networks develop under different conditions than elsewhere, and our goal is to provide data for studying this phenomenon.

In this dataset paper, we present a data collection of a national Twittersphere in a West African country of conflict. While not the largest social network in terms of users, Twitter is an important platform where people engage in public discussion. The focus is on Mali, a country beset by conflict since 2012 that has recently had a relatively precarious media ecology. The dataset consists of tweets and Twitter users in Mali and was collected in June 2022, when the Malian conflict became more violent internally both towards external and international actors. In a preliminary analysis, we assume that the conflictual context influences how people access social media and, therefore, the shape of the Twittersphere and its characteristics. The aim of this paper is to primarily invite researchers from various disciplines including complex networks and social sciences scholars to explore the data at hand further. We collected the dataset using a scraping strategy of the follower network and the identification of characteristics of a Malian Twitter user. The given snapshot of the Malian Twitter follower network contains around seven million accounts, of which 56,000 are clearly identifiable as Malian. In addition, we present the tweets. The dataset is available at: https://osf.io/mj2qt/

## INTRODUCTION

First and foremost, this article presents and offers a preliminary analysis of the Twittersphere of Mali, a landlocked West African nation that shares a border with Senegal, Mauritania, Ivory Coast, Niger, Burkina Faso, and Algeria. The construction of the Malian Twittersphere is part of the research project *Decoding Digital Media in African Regions of Conflict* [DD-MAC], which seeks to study the role social media usage plays in mediating conflict in African countries. The project members compare social media and conflict dynamics in Ethiopia and Mali. It builds on Philipp Budka and Birgit Bräuchler (2020)'s suggestion that information flows play a crucial role in hybrid war countries. The article and its dataset intend to serve as a starting point for researchers interested in unraveling social media's characteristics in vulnerable media environments.

Since 2009, Malians have increased access to social media, partially as a consequence of Facebook policies [Keita et al. 2015; Nothias 2020]. Since then, Mali has become a networked public [boyd 2010] with Facebook and WhatsApp being especially popular communication channels. However, we decided to concentrate on Twitter because participants in the discussion fora we held in Mali in May 2022 indicated that discussions and interactions in the Twittersphere relate to political and social developments in the country [Mutsvairo et al. 2023]. By Malian Twittersphere we refer to Twitter users located in Mali and those that express a strong connection to Mali, which is often the case for the Malian diaspora. Twitter is an essential platform for the release of journalistic and activist information. Malians move between Twitter and platforms such as Facebook and TikTok to communicate about particular issues of their contemporary world. Mirca Madianou and Daniel Miller [2013] argue that the choice of a medium has social and moral consequences as it is embedded in interpersonal relationships. They refer to the interlinked structure of individual media as polymedia. Therefore, the Twittersphere can be that space where people create, maintain, and shape discussions on politics, social issues, and health.

Additionally, pragmatic reasons motivated us to collect Twitter data in the first instance, as Twitter's API allowed researchers to simply collect user activity data. Until May 2023, researchers could access large amounts of tweets and user data from Twitter for free. Under these circumstances, we collected the dataset from February 4th 2022 until June 24th 2022. Currently, researchers can access only small amounts of Twitter data, which decreases scholars' opportunities to analyze Twitter communication. The access to small amounts of data is especially relevant for studies where a large number of accounts must be collected before accounts of interest can be identified, as is the case in our mapping of a national Twittersphere. With this study, we also sought to better understand Twitter and its usage in the Global South and contribute knowledge needed in foregrounding Digital Humanities among African scholars [Ngué Um and Jones 2022]. Against this backdrop, the presented database is an important contribution to scholarship on Twitter as it shares a historical snapshot of the size of a Twitter user base in Africa. Scholars can use the data for multiple qualitative purposes, such as comparative analyses with national Twitterspheres in West Africa or the Global North [Schroeder et al. 2022], as well as in-depth case studies of linguistic, topical, and event-based communication on Twitter by selected Malian accounts.

We contribute to the few existing datasets of African countries that provide social media posts for amongst others sentiment analysis and Natural Language Processing (NLP) [Cornelissen et. al 2019, Muhammed et. al 2023]. In doing so, we aim to address the geopolitics of knowledge on digitalization and the imbalance between research on Twitter in the Global North and South in particular. Furthermore, we provide a preliminary analysis of the dataset and answer the following questions: what is the scope of the Malian Twittersphere? What are the specificities of the Malian Twitter network in relation to its functionality in the media ecology of the country?

**I Twitter reserch in the Global South**
Most of the Twitter researchers that apply digital tools and methods have concentrated on societies and events in the Global North, for example by looking at and visualizing topic and user networks [Chiluwa 2015, Gelfgren 2016, Logan et. al 2023]. In Africa, Twitter research is primarily qualitative and concentrated predominantly on South- and West Africa, and mainly countries where the official language is English like Nigeria, Kenya, and Ghana, with the exception of Senegal [Aduloju 2016, Shipley 2017, Egbunike 2018, Ezeh & Mboso 2018, Jordan 2018, Ndiaye 2021]. These qualitative studies illustrate the growth of social media usage, the content, and who the actors are. A frequently used method in such research is hashtag studies by looking at content related to instant upheavals, activism, and revolts [see for instance examples on South Africa: Munoriyarwa et al. 2022; Smit and Bosch 2022, Bosch 2017; West Africa: Roy et al. 2020; Nigeria: Egbunike et al. 2015; and the Arab Spring: Bruns et al. 2013]. Additionally, some scholars work with Twitter corpora to do computational network analysis, such as Stefanie Stratchan and Arurona Gerber [2019], who studied the network structures in South African Twitter communities. Furthermore, Laurenz Cornelissen et al. [2019] analyzed sentiment and collected data of political and social events on Twitter in South Africa to detect and define communities aligned with particular thematic issues using community detection. We did not find a study consisting of a holistic dataset of Twitter in a particular country in conflict. Hence, this paper presents a dataset to compare and contribute to previous findings on Twitter usage in the Global South.

As Barbara Poblete et al. [2011] show with a comparison of Twitterspheres in ten countries, Twitter networks differ by national context. One finding is that reciprocity, which is generally low in Twitter networks, would change when networks are smaller. The authors generally agree that Twitter functions as a 'news' provider, where the few people who 'produce' the news have many followers [Ediger et al. 2010]. Axel Bruns and Jean Burgess [2012, 804] defined Twitter as "...both a social network site and an ambient information stream." Twitter is a medium where people participate to access information on recent developments or societal problems. Therefore, Twitter is especially functional for people in countries or regions where public media are undeveloped, hard to access, or controlled by the state or an oligarchy. In earlier research on Mali, we show that Malians partially use social media because it gives them a form of freedom of press, which is limited in government-run public media. Moreover, in the present-day Malian media ecology social media usage is dominant compared to the older media, while journalists and citizens generate most information from social media sources and through word-of-mouth [see Mutsvairo et al. 2023]. Hence, understanding the Twitter network in Mali is essential to knowing who circulates what information.

**II Capturing the Malian Twittersphere**
Twitter is a dynamic social network, meaning relationships form and disperse over time [Carley 2003; Sekara et al. 2016; Trémolières et al. 2021]. While individual actors make up the networks underlying Twitter, it is the interdependent and interpersonal structure between active users that makes the network meaningful [Yang and Sun 2021]. Twitter is a genre of "networked publics" [boyd 2010]. This means that networked technologies structure the dynamic ways in which people interact and debate in this online environment. While the dynamic nature of the Twitter network is a fascinating aspect to study, as interactions between people are constantly changing and people delete and suspend accounts, we cannot continuously update the dataset and therefore present a static snapshot of the Malian Twittersphere in a particular period of conflict. The snapshot is a starting point for analyses of Twitter users' cultural, linguistic, and temporal interactions

To generate a representation of the Malian Twittersphere we used a mixed-methods approach. Researchers have studied Twitter using such approaches, among them digital humanities [Bruns et. al 2014], computational ethnography [Breslin et al. 2020] and machine anthropology [Pretnar 2020]. These methods mix computation with ethnographic, literary, or historical observation and focus on social interactions in physical or online environments through domain expertise. Here, domain expertise refers to in-depth knowledge of a particular field that researchers can, for example, acquire through ethnographic fieldwork or sociological research. Domain expertise is an essential prerequisite for the evaluation of computer analysis and the development of a comprehensive data mining strategy. Data mining involves the identification of patterns and trends in the data, for which success depends on finding appropriate filters that allow the determination of an individual's affiliation with a target population (see Section IV). Furthermore, domain expertise is necessary for validation [McCue 2014], because it enables researchers to recognize local patterns and identify limitations to the validity and reliability of the computational data.

Our approach to capturing the Malian Twittersphere consists of two phases:
1. We first develop a scraping strategy based on Twitter's follower network.
2. Then, we find characteristics with which an individual user can be identified as Malian.

In the following, we introduce each of the two phases in detail.

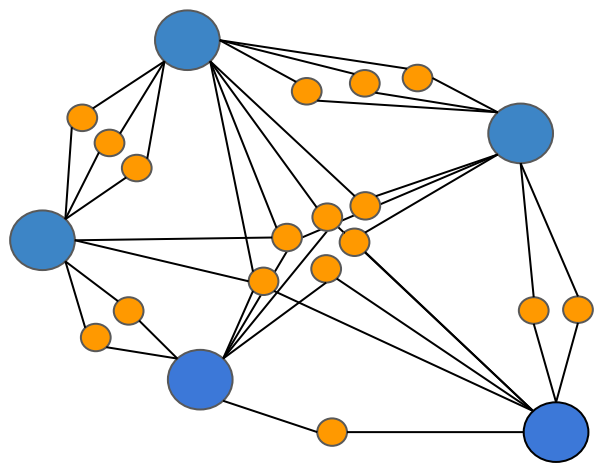

Figure 1. A follower network. In blue are the Twitter accounts we collected manually and in orange the followers between them.

**III Collecting the Malian Twittersphere based on the Twitter follower network**
To grasp the Malian Twitter network, we first hand selected well-known Malian Twitter accounts that tweet frequently about societal issues starting February 4th 2022 (see Appendix 9.1). This initial set of accounts consists of a mix of governmental institutions, politicians, news agencies, and bloggers. To identify these figures we looked at follower count and used domain expertise to identify users who frequently tweet on Mali. However, there are various methods to determine influential users in a network. For example, through a social network analysis of Twitter interactions Austin Logan et. al [2023] aimed to identify topically-focused groups and used the HITS algorithm.

Journal of Data Mining and Digital Humanities                                         http://jdmdh.episciences.org
ISSN 2416-5999, an open-access journal

After selecting these accounts, we proceeded iteratively and collected the friend- (people who mutually follow one another) and follower networks of these users to enlarge our data set manually. The manual enlargement process took place between February 4th 2022 and June 22nd 2022. The assumption is that such accounts are likely to be part of the Malian Twittersphere. Friend- and following lists are relevant data sets on social network sites, because they are an "articulation of a public" [boyd 2010: 5] and thus represent networked publics. The friends and following lists give an impression of who constitutes the social networks of the initially selected users.

To discover the entire Malian Twitter network, let $G = (A, E)$ be Twitter's entire follower graph. We proceeded iteratively from the hand-selected set of well-known Malian Twitter accounts (see Appendix 9.1). Let $U_1$ be the initial set of users in iteration $t = 1$. We take each user $u \in U_t$ and collect its followers $F_u$. Then we build the follower network $G_u = (\{u\} \cup F_u, E_u)$ for each user $u$. Here, $E_u$ is simply a set containing one directed edge $(u, f)$ for each $f \in F_u$. Let $F_t = \cup_u F_u$ be the set of all followers in iteration $t$.

We select the next set of users $U_{t+1}$ by manually checking the most promising candidates among the followers in $F_t$. To find the most promising candidates, we defined the degree of a follower $f \in F_u$ (Represented by orange dots in Figure 1) in iteration $t$ as $deg(f) = |\{(u, f) \in E | u \in U_t\}|$. As a criterion for affiliation, the accounts were then manually assigned. For this purpose, we took followers and friends, the content of tweets and retweets, and the profile properties into account. Additionally, we defined the number of followers a follower $f$ has as *followers(f)*. Please note that this follower count is the number of followers indicated by a particular Twitter user profile.

We sorted the followers by degree, breaking ties by their follower count and manually checked the followers with the highest degree until we obtained a set $U_{t+1}$ of 40 accounts verified to be Malian. We chose this number based on the available human annotators. We then started the next iteration. Figure 1 illustrates the process described above. The dark blue nodes represent the already labeled users. The orange nodes represent the immediate followers that will be labeled in the next round.

We repeated this process until we obtained a set of 400 Malian users by June 22nd 2022. Afterward, we replaced the manual selection process with automated filtering running the entire data collection that led to the final data set between June 22nd and June 24th 2022. Although, for the first iterations, we dealt with a selection bias from the initial sets $U_t$, we argue that repeating that process and manually selecting the top $N$ accounts is sufficient to capture a solid user base. Furthermore, we argue that the proposed collection method results in sufficient generalizability because we can assume that we are working with a small-world network [Watts and Strogatz 1998]. Nevertheless, this corpus does not aim to be comprehensive, rather it offers a large segment of the Malian population.

Since our initial set consisted of accounts with many followers, manually checking all these followers was not possible. Sampling the set of followers revealed that it contained many accounts that were neither Malian nor part of the diaspora. Hence we had to use additional automated filtering before the manual inspection could be applied. We did this by looking at profile characteristics described below. Eventually, we selected the final set of accounts based on a dual observation strategy on June 22nd 2022: we obtained Twitter profiles, selected a random sample of 300 profiles, and checked for errors. We repeated this process while refining the filter characteristics until the sample contained few or no non-Malian accounts. As a result, we ended up with a corpus containing only a small number of user profiles that cannot be assigned to the Malian Twittersphere with certainty. These accounts are news agencies that mention one of the properties in their profile.

## IV Characteristics of a Malian Twitter user

It is not easy to define the national and geographic origins of a Twitter account, because users have various strategies to represent themselves. Twitter users can choose to show their account location and tweet location. An account location may or may not contain geolocation. This characteristic is ambiguous because people do not always fill in correct and precise locations. If enabled, a tweet location contains geolocation coordinates, which based on GPS shows the location someone tweeted from. Yet, this characteristic does not clarify where the person lives. Twitter is a platform that does not provide geographical information or requires its users to provide a true indication of their place of residence. Therefore, it is necessary to infer belonging to the Malian population from alternative, user-provided information. In a study on the Australian Twittersphere, scholars for example looked at particularly national hashtag topics, timezone information in user profiles, and additional snowball crawl of the follower and friend lists [Bruns et. al 2014, 118]. We started with the snowball crawl and continued focusing on the self-reported properties of users.

Although many user profiles show self-reported properties that allow for an assignment to a particular nationality, such as location or profile description, there are cases in which the affiliation is less obvious. Unlike some countries with a national language unique to that country, the official language in Mali is French. Selecting users who use the French language is at best meta-information that helps to constrain the search. However, it will not filter out Malian users in particular, as French is used globally across national populations. Moreover, filtering accounts based on self-reported features risks the inclusion of inauthentic accounts who claim to be part of a particular national twitter community while being controlled from a foreign country. Research shows such imposter accounts occur in places of conflict[1]. While we did not expect there to be many imposter accounts in the dataset, we performed a manual check in which we indeed did not find any suspicious accounts. The manual check consisted of visiting Twitter profiles and labeling the accounts we collected through friend- and follower networks. Due to the manual exploration during these first iterations (see Section III), we also observed where, how, and who uses particular identification features. This process helped us to identify the characteristics of Malian Twitter profiles.

To determine whether a user profile belongs to a Malian, we considered a combination of profile properties and properties of the profile's follower network. Each Twitter user profile contains a user name, a self-provided user description, and an optional location field. Based on these features, we developed a list that includes the most frequently reported locations (see Appendix 9.2), frequently used words (see Appendix 9.3), and emoticons (see Appendix 9.4). This filtering process that includes location and description fields to identify national Twitter users has been applied elsewhere to collect the Norwegian Twittersphere [Bruns & Enli 2018]. We assigned a user to the Malian Twittersphere if the person reported one of the locations or a combination of features in their self-reported location, username, or profile description. Developing this list was an iterative process that involved our evaluation of mentioned characteristics. For example, city names occurred in different spellings, and some users reported regions, parts of cities, or village names. In addition to location, there were terms that Malians frequently used in profile descriptions, such as the words *malien, malienne, and malikura* or the country code of Mali, +223. Furthermore, the Malian flag was a popular symbol among Malian Twitter users. The variety of these features shows that determining the nationality of a Twitter user requires domain expertise and continuous checking for accuracy.

**V The features of the dataset**

---
[1] https://blog.twitter.com/en_us/topics/company/2021/disclosing-state-linked-information-operations-we-ve-removed

Using the presented method, we collected 56,505 Malian Twitter accounts along with 10,676,046 tweets and retweets from these accounts. Existing estimations of the Malian offline-, Twitter-, and Internet population help to assess the reliability of our dataset. According to International Telecommunication Union (ITU) statistics 68% of the Malian population had access to 3G networks in 2021. According to Internet World Statistics (IWS) out of a population of $20.86M$ in 2022, $12.48M$ people had access to the Internet in December 2021, this accounts to 59.8% of the total population. According to DataReportal there were $6.33M$ Internet users in Mali in January 2022, which is 29.9% of the total population. These user figures diverge but give an indication that at least 50% of the Malian population has access to mobile Internet and can access social media. Additionally, DataReportal's figures show there were 56.200 Twitter users in Mali in January 2022, which is similar to our estimate in June 2022.

In addition to the accounts we identify as Malian, using the iterative addition of followers, we found 7,025,250 accounts that do not match the filter criteria and thus we do not consider them to be Malian. While these accounts may be part of the Malian Twittersphere, they are not identifiable as such. However, as depicted in Figure 1, we refer to these accounts as the outskirts of the follower network. Based on our analysis we also provide the degree distribution of the follower network subdivided into the central and entire network in Figure 2 (left).

Besides the Twitter IDs for tweets and user accounts including a corresponding Python hydrate script, we publish the entire follower network. For the latter, to be compliant with the Twitter Terms of Service, we anonymized the user IDs by replacing them with consecutive numbers starting from 0. However, we provide a file in which we resolve the mapping of anonymized ids to tweet ids. Furthermore, we mark each anonymized user id that we identified as Malian.

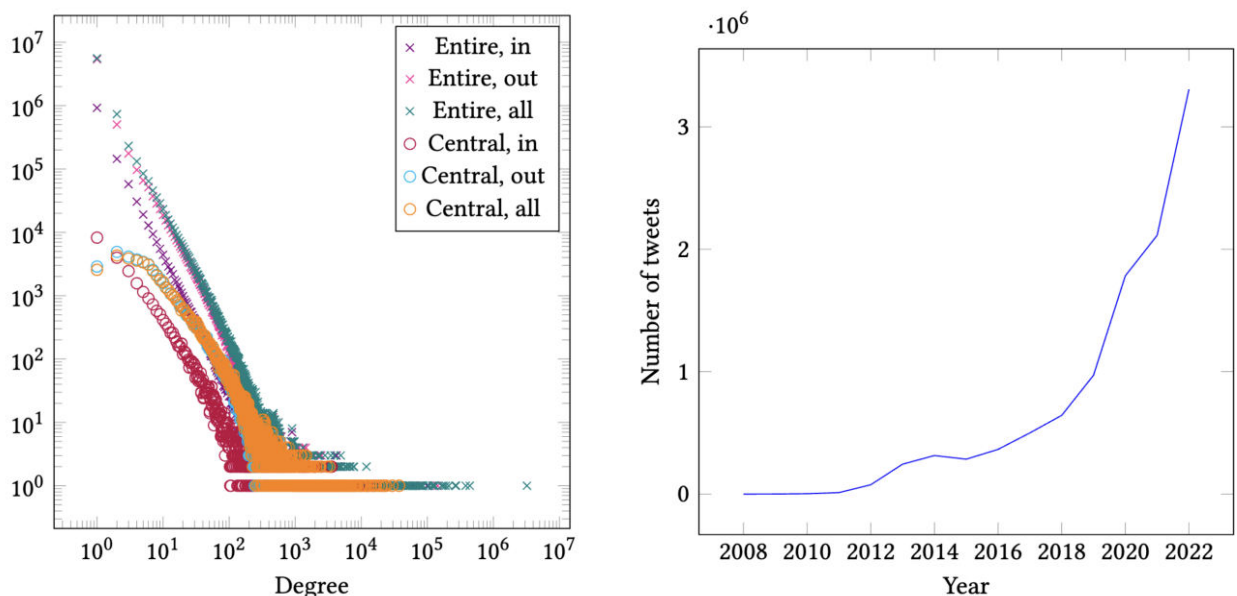

Figure 2. (Left) Comparison of the degrees of the entire network and the central network. The crosses represent the entire network, while the circles represent the central network. (Right) The distribution of the number of tweets per year.

The presented dataset consists of Twitter users who fitted the Mali "label" and whom we thus identified as members of the community. This identification results in a bias, as the dataset lacks a potential group of users of the Malian Twitter community who use different national identification strategies. Nevertheless, the corpus of the given period is representative and can be used to analyze popular content in the network and the geographic clustering of Twitter users.

**VI Preliminary analysis of the dataset**

The data shows Twitter usage is growing and social media is gaining importance. This finding correlates with our previous finding that social media has largely taken the place of traditional media in Mali as the primary source of information. The Twittersphere as we captured it, is set in a particular historical timeframe. 2022 was a year of societal disruption, hence scholars should interpret and analyze the tweets accordingly. Figure 2 (right) shows the growth of the number of tweets on a year-by-year basis. The figure shows that Twitter adoption in Mali is recent and growing fast, with people tweeting far more in 2020, 2021, and 2022 than in the preceding years. This development coincides with the growth of conflict incidents in Mali. The country has been subject to an intersection of armed conflicts, often between different ethnic and religious groups since 2012. Various armed groups strive for power, land, and resources. Every year insurgencies increase and meanwhile the coup in May 2021 by Colonel Assimi Goïta has consolidated the army's power and state repression. While Malians frequently make social media comments on politics, ethnicity, and violent militias, interviewees in previous research taught us that increasingly bloggers, influencers, and local social media outlets play an authoritative role in the dissemination of news. Questions around bias, truth, and ownership of information are important to understand the current conflict in Mali and neighboring countries with similar insurgencies such as Burkina Faso. The presented dataset can give insight into these issues and help assess West-Africa specific characteristics of Twitter usage in a period of conflict. We plan to address and analyze the relationship between Twitter growth and content on conflict in future articles, as well as which actors are part of the presented network.

Figure 3 (left) shows that the average number of tweets is large in big urban centers, such as Bamako, Sikasso, Mopti, Kayes, and Segou. Simultaneously, the plots show that users in cities with smaller populations, such as Gao, Koutiala, and Tombouctou, are equally or even more active. This finding may lead to the conclusion that individuals in large urban centers, where education levels are generally higher, are more likely to use Twitter. Moroever, this finding is in line with informants' narration that Twitter unites elites. Moreover, most of these urban centers are in regions of intense conflict. Thus, the numbers indicate that high Twitter activity coincides with the intensity of conflict. Our previous research indicates that regional and local content is often on violence. Hence, the dominance of users from conflict-struck zones in this dataset may point us to such thematic content.

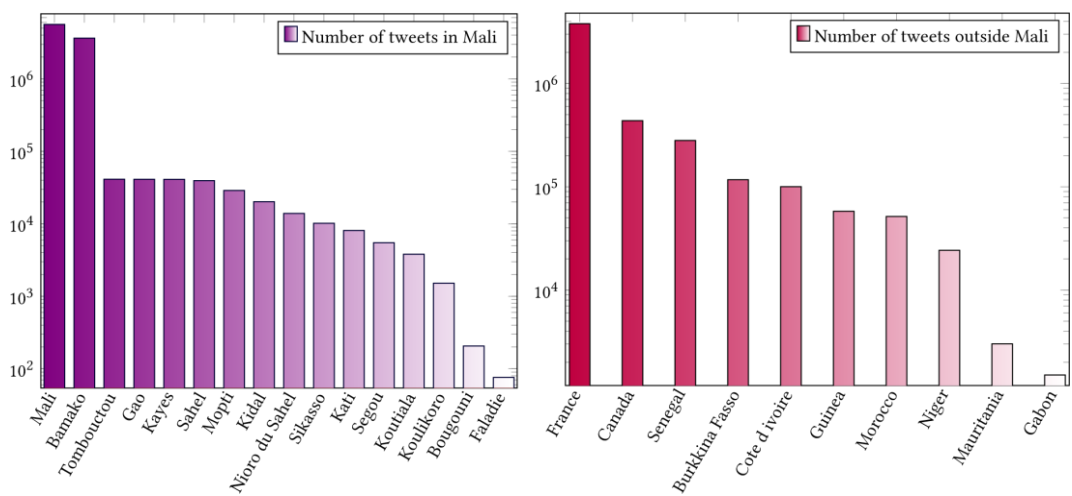

Figure 3. Origin of Tweets in the Malian Twittersphere. (Left) Number of tweets in large urban centers in Mali in our dataset. (Right) Number of tweets by members of the Malian diaspora by country of residence.

The manual checks of Twitter accounts also confirm this hypothesis, as we learned that Twitter users often address issues regarding the violent conflict in Mali. We also noticed that politicians and bloggers who discuss conflict usually have many followers and tweet on a daily or weekly basis. They thus have an essential role in the Twittersphere. These findings indicate that the presented dataset is suited for studying the conflict, which is why we aim to bring it to the attention of scholars in the field. In-depth or comparative analysis of the actors in the dataset to other datasets would help to understand the relationship between leadership, conflict, and digitalization.

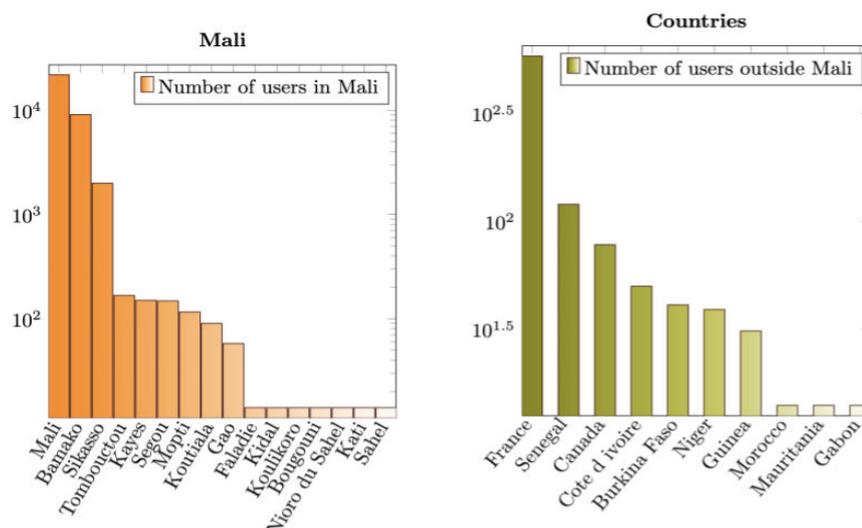

Figure 4. Distribution of users in the Malian Twittersphere. (Left) Number of users in large urban centers in Mali in our dataset. (Right) Number of users in the Malian diaspora by country of residence.

Figure 4 (right) shows that the number of users outside of Mali is small. The majority of users outside Mali come from France, Francophone Canada, and countries across West-Africa. But while their number is small, the activity of the user from these regions is high, as shown in Figure 3. For example, the number of tweets from users in France is comparable to the number of tweets from Bamako, despite the fact that the number of users in Bamako is more than 20 times higher. A detailed analysis of the diasporan and foreign participation in the Malian Twittersphere would fill the research gap on the digital role of diaspora groups in homeland conflicts.

**VII Conclusion**

With this paper, we present a snapshot of the Malian Twittersphere and invite other researchers to participate in investigating the reciprocal effects between social media usage in Mali and societal discussions. Researchers can for example do such research by looking at the content of tweets through topic modeling or discourse analysis. The preliminary analysis of the dataset shows that Twitter is primarily used in large cities, the platform has a growing number of users, and people based outside of Mali form a significant group of tweet producers. Although only a relatively small percentage of the total population of Mali is active on Twitter, many of the main contributors are public figures or politicians. The impact of these actors and their communication means should not be underestimated in the authoritarian context of Mali. Hence, media scholars and computational scientists can use this dataset as a starting point to reflect on the politicization of West African media spaces and compare the Malian Twittersphere to other global Twitterspheres. Moreover, this corpus is a contribution to digital archives on Mali, which can help build a historical social media record.

# IX Appendix
## 9.1 Hand-selected set of initial accounts

| Blogger | Username | Followers | Friends |
|---|---|---|---|
| 1 | xxxxx | 25.4004 | 5.237 |
| 2 | xxxxx | 12.700 | 3.278 |
| 3 | xxxxx | 7.910 | 1.062 |
| 4 | xxxxx | 6.894 | 475 |
| 5 | xxxxx | 2.844 | 4.555 |
| 6 | xxxxx | 185 | 8 |
| 7 | xxxxx | 140 | 160 |
| 8 | xxxxx | Private | Private |
| 9 | xxxxx | Deleted | Deleted |

Table 1. Initial set of users ordered by number of followers.

## 9.2 Locations characteristic for Malian Twitter users

bamako, #bamako, kayes, #kayes", koutiala #koutiala, kati, #kati, korofina,#korofina", sikasso, #sikasso, manankoro, #manankoro, koulikoro, #koulikoro, baguineda, #baguineda, kolokani, #kolokani, segou, #segou, tombouctou, #tombouctou, mopti, #mopti, faladie, #faladie, kidal, #kidal, sebenikoro, #sebenikoro, ségou, #ségou, niamakoro, #niamakoro, timbuktu, #timbuktu, sahel, #sahel, nioro, #nioro, bougouni, #bougouni, kalabancoro, #kalabancoro, kalaban, #kalaban, kolondieba, #kolondieba, niono, #niono, koumantou, #koumantou, azawad, #azawad, sokorodji, #sokorodji, gao, #gao, nioro, #nioro|, kadiolo, #kadiolo, djikoroni, #djikoroni, massigui, #massigui, sébénikoro, #sébénikoro, faladiè, #faladiè, sanankoroba, #sanankoroba, dialakorobougou, #dialakorobougou, sikoro, #sikoro, quelessebougou, #quelessebougou, dialakorodji, #dialakorodji, wassoulou, #wassoulou, fulani, #fulani, sirakoroba, #sirakoroba, sirakoro, #sirakoro, kalaban, #kalaban, menaka, #menaka, sikoroni, #sikoroni, djalakorodji, #djalakorodji, torokorobougou, #torokorobougou, tabakoro, #tabakoro, tonka, #tonka, korodougou, #korodougou, ménaka, #ménaka, diallassagou, #diallassagou, nènèkoro, #nènèkoro, royaume, #royaume, kalabankoro, #kalabanko

## 9.3 Terms characteristic for Malian Twitter users
mali, #mali, malien, #malien, malienne, #malienne, malikura #malikura, malian, #malian, beautifulmali, #beautifulmali, +223, # + 223, madeinmali, #madeinmali, 00223, #00223

## 9.4 Emoticons characteristic for Malian Twitter users